\documentclass[11pt,twoside]{article}
\usepackage{macro-fsut-eng}
\usepackage{psfig}
\usepackage{graphicx}

\usepackage[T1]{fontenc} 

\usepackage{latexsym}
\usepackage{verbatim}

\begin{document}

\vskip 1.0cm
\markboth{J.E.~Forero-Romero et al.}{Panchromatic view of high-z galaxies}
\pagestyle{myheadings}

\vspace*{0.5cm}
\title{Towards a panchromatic picture of galaxy evolution during the reionization epoch}

\author{Jaime E.~Forero-Romero$^1$,  Gustavo Yepes$^2$, 
  Stefan Gottlöber$^3$ \\
  and Francisco Prada$^{4,5,6}$\\}
\affil{
$^1$Department of Astronomy, University of California, Berkeley, CA 94720-3411, USA\\
$^2$Grupo de Astrof\'{\i}sica, Universidad Autónoma de Madrid,   Madrid
E-28049, Spain\\
$^3$Leibniz-Institut für Astrophysik Potsdam (AIP), An der Sternwarte 16, 14482 Potsdam, Germany\\ 
$^4$Campus of International Excellence UAM+CSIC, Cantoblanco, E-28049 Madrid, Spain\\
$^5$Instituto de Física Teórica, (UAM/CSIC), Universidad Autónoma de Madrid, Cantoblanco, E-28049 Madrid, Spain\\
$^6$Instituto de Astrof\'{\i}sica de Andaluc\'{\i}a (CSIC), Glorieta de la Astronom\'{\i}a, E-18008 Granada, Spain
}

\begin{abstract}
  There are thousands of confirmed detections of star forming galaxies
  at high redshift $(z>4)$. These observations rely primarily on the
  detection of the spectral Lyman Break and the Lyman$\alpha$ emission
  line. Theoretical modelling of these sources helps to interpret the
  observations  in the framework of the
  standard cosmological paradigm.  We present results from the {\em High-z
  MareNostrum Project}, aimed at constructing a panchromatic picture of
  the high redshift galaxy evolution that will improve our understanding of young
  star forming galaxies. Our simulation successfully reproduces the
  observational constraints from Lyman Break Galaxies and
  Lyman$\alpha$ emitters  at   $5<z<7$ . Based on this model we make  predictions on
   the expected Far Infrared  (FIR)  emission that should be observed for LAEs.
  These predictions will help to settle down the question on the dust content
  of massive high-z galaxies, an issue that will be feasible to probe
  observationally with the Atacama Large Millimetre Array (ALMA).
\end{abstract}

\section{Introduction}

Observationally, there are two main ways to detect high redshift star
forming galaxies. The first technique uses broad band measurements in
order to detect the drop in flux due to the Lyman Break in the  galaxy's
Spectral Energy Distribution (SED). Galaxies detected through this
technique receive the name of Lyman Break Galaxies (LBGs). The second
technique aims at detecting the Lyman$\alpha$ emission line either by narrow
band filtering techniques or by direct spectral measurements. Galaxies
detected through this selection technique receive the name of Lyman$\alpha$
Emitters (LAEs).  For some of the high-z LBGs there has also been 
detection in their rest-frame near-IR which has allowed to provide an
estimation of their stellar masses.

In this work we describe our efforts   to build a panchromatic picture
of high-z galaxies at  $5<z<7$. Our work is based on the results from 
a large hydrodynamics simulation that follows the gravitational collapse of
dark matter and gas  and including  radiative processes and the
ensuing star formation, together with associated chemical and
mechanical feedback from exploding stars.  Based on the analysis of the
numerical galaxies, we have suggested a  model
to account for the dust extinction and the radiative transfer of the Lyman$\alpha$ line. We
have shown  how this model is in fair agreement with the observed
properties of LBGs and LAEs (Forero-Romero et al. 2010,2011). Special attention was made to
reproduce  the observed fraction of LBGs with a strong Lyman$\alpha$
emission (Forero-Romero et al. 2012). In what follows, we will describe the main features of our
model and will conclude  by pointing out the expected trends for the
FIR emission that will  be probed by ALMA  in the near future.

\section{The Simulation and Galaxy Finding}
\label{sec:sim}
The High-Z MareNostrum Simulation follows the nonlinear evolution of a
cubic region of $50$ $h^{-1}$ Mpc (comoving) on a side. The dark matter and gas
distributions are sampled with $1024^3$ particles for each component. 
The  gas physics  includes radiative cooling and photoionisation from
an  homogeneous UV background switched on at $z=6$.
 Star formation and feedback (chemical and dynamical) are
included following  the Springel and Hernquist (2003) model. We identified
all gravitationally  bound objects using the Amiga Halo Finder
(Knollman \& Knebe 2009). All objects with  at least 1000 particles
are kept for further analysis.  More details can be found  in
Forero-Romero et al. (2010).

\section{Modelling the UV Continuum and the Lyman$\alpha$ line}

The stellar SED is constructed from the stellar particles contained in
the galaxies detected as described in Section \ref{sec:sim}. The main
assumption under this calculation is that each stellar particle in the
simulation can be treated as a burst of star formation with a given
mass, age and  metallicity. We assume a constant Salpeter IMF.
  
We implement a phenomenological model to quantify the extinction
produced by dust in the stellar SED. The physical model assumes that
extinction is  divided  into two contributions that affect different
stellar populations of different ages. Old stellar populations see an
effective extinction from an homogeneous Interstellar Medium (ISM),
while young stellar populations suffer additional extinction due to the
molecular clouds where they are embedded at birth.  

The calculation of the escape fraction of Lyman$\alpha$ radiation assumes
the same geometry as the UV continuum extinction.  We have developed
the Monte-Carlo radiative transfer code CLARA to follow the path of
Lyman$\alpha$ photons through the ISM in simplified geometrical
configurations. Details of  the implementation of this model can be
found elsewhere (Forero-Romero et al. 2011).

\section{Results}
\begin{itemize}
\item {\bf LBGs, LAEs and Stellar Masses}:
Our numerical  results for the luminosity functions for LBGs are in good agreement with the observational results. Only a minor disagreement is noticeable at the faint end of the luminosity function, where there
seems to be an overabundance of simulated dwarfs with respect to the
observations.  UV colours as parametrised by the $\beta$ slope as a
function of restframe UV magnitudes are also in agreement with
results derived from observations. This shows that the dust correction
we have applied to the simulated galaxies is reasonable (Forero-Romero
et al. 2010).  The simulated LF of LAEs  shows a good agreement with observations at the bright
end, while showing an overabundance at the faint end, which is somewhat
more noticeable than in the restframe UV (Forero-Romero et al. 2011). Recently, we have shown that the  stellar masses as a function
of intrinsic UV luminosity are also in good agreement with the
observational estimates, at least for the most massive systems
(Forero-Romero et al. 2012b).

\item{\bf LBGs as LAEs}:
As part of the   international  observational effort to probe the
reionisation epoch  using LAEs and LBGs, a
new kind of observations were conducted. Galaxies detected primarily as
LBGs were then followed up spectroscopically. From these measurements
one can determine whether a LBG is also a LAE by requiring the
equivalent width to be larger than a fixed threshold value. Using these
results  we can consider now a new kind of statistics, the fraction of LBGs
showing strong Lyman$\alpha$ emission, $X_{{\rm Ly}\alpha}$.  The results from our simulations, without any further change, fully
agree  with the observational measurements of this fraction. We
have also shown that  the Lyman$\alpha$ escape fraction
is decreasing with galaxy mass (Forero-Romero et al. 2012).

\item {\bf FIR Emission}
For each galaxy in the simulation we calculate the expected flux
measured in the ALMA band at 353 GHz as a function of observed Lyman$\alpha$
luminosity. We find that the brightest LAEs do not have a high FIR
luminosity, passing completely undetected by ALMA (Fig. 1). This prediction of
our model can be quantitatively understood as follows. Bright galaxies
in the FIR have a high star formation rate and dust mass contents. These
high dust values and neutral hydrogen associated with the most massive
 FIR bright galaxies naturally corresponds to very low escape
fractions. 

\end{itemize}

\begin{figure}  
\begin{center}
\hspace{0.25cm}
   \psfig{figure=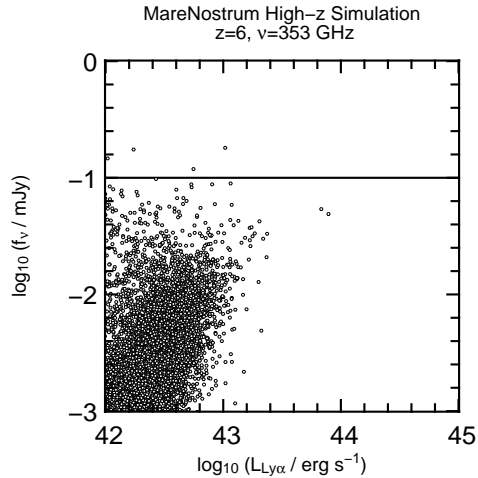,width=6.cm}
\caption{Flux in the ALMA band at 353 GHz as a function of the
  observed Lyman$\alpha$ emission. The horizontal line represents the
  threshold for detection. In our model the brightest Lyman$\alpha$ galaxies won't
  be detected by ALMA.}
\label{alma}
\end{center}
\end{figure}

\section{Conclusions and Outlook}
We have constructed a model for high redshift galaxies (LBGs and LAEs)
based on the results of the {\em High-z MareNostrum Simulation}.
 From the numerical results we are able to estimate the observed
 restframe UV and Lyman$\alpha$ emission in these galaxies.

We find that our results are  broadly consistent with the current
observational constraints. The largest discrepancy  is an 
overabundance in the faint end LF both for LBGs and LAEs. The agreement
between our model and observations for objects brighter than $M_{\rm
  UV}<-20$ is remarkable. 

We also  predict the expected flux at 353 GHz as a function
of the observed Lyman-$\alpha$ emission. In  our model, the brightest LAE
galaxies are too  faint FIR emitters. Correspondingly,
bright IR sources will be detected by ALMA, but they will not be detected
as bright LAEs.

\section*{Acknowledgements}
J.E. F-R is supported by the Peter and Patricia Gruber Foundation,
through their fellowship administered by the International
Astronomical Union. 
We thank the Barcelona Supercomputing Center  for giving us access to
the Marenostrum Supercomputer. We also thank MICINN (Spain) for
financial support  under projects AYA2009-13875-C03-02 and
FPA2009-08958.

\end{document}